\makeatletter
\@ifundefined{@parse@version@dash}{%
\def\@parse@version#1{\@parse@version@0#1}
\def\@parse@version@#1/#2/#3#4#5\@nil{%
\@parse@version@dash#1-#2-#3#4\@nil}
\def\@parse@version@dash#1-#2-#3#4#5\@nil{%
  \if\relax#2\relax\else#1\fi#2#3#4 }
}{}
\makeatother

\documentclass[aip,reprint]{revtex4-2}
\usepackage{bm}
\usepackage{color}
\usepackage{graphicx}
\usepackage{lineno}

\begin{document}

\title{Robust perpendicular magnetic anisotropy in Ce substituted yttrium iron garnet epitaxial thin films}

\author{Manik Kuila}
\affiliation{UGC-DAE Consortium for Scientific Research, University Campus, Khandwa Road, Indore 452001, India.}

\author{Archna Sagdeo}
\affiliation{Synchrotrons Utilization Section, RRCAT, Indore India.}
\affiliation{Homi Bhabha National Institute, Training School Complex, Anushakti Nagar, Mumbai 400094, India.}

\author{Lanuakum A Longchar}
\affiliation{School of Physics, University of Hyderabad, Hyderabad-500046, India}

\author{R.J.Choudhary}
\affiliation{UGC-DAE Consortium for Scientific Research, University Campus, Khandwa Road, Indore 452001, India.}

\author{S.Srinath}
\affiliation{School of Physics, University of Hyderabad, Hyderabad-500046, India}

\author{V.Raghavendra Reddy}
\email{varimalla@yahoo.com; vrreddy@csr.res.in}
\affiliation{UGC-DAE Consortium for Scientific Research,  University Campus, Khandwa Road, Indore 452001, India.}

\begin{abstract}

Cerium substituted yttrium iron garnet (Ce:YIG) epitaxial thin films are prepared on gadolinium gallium garnet (GGG) substrate with pulsed laser deposition (PLD). It is observed that the films grown on GGG(111) substrate exhibit perpendicular magnetic anisotropy (PMA) as compared to films grown on GGG(100) substrate. The developed PMA is confirmed from magneto-optical Kerr effect, bulk magnetization and ferromagnetic resonance measurements. Further, the magnetic bubble domains are observed in the films exhibiting PMA. The observations are explained in terms of the growth direction of Ce:YIG films and the interplay of various magnetic anisotropy terms. The observed PMA is found to be tunable with thickness of the film and a remarkable temperature stability of the PMA is observed in all the studied films of Ce:YIG deposited on GGG(111) substrate. 

\end{abstract}


\maketitle \section{Introduction}

Cerium substitute yttrium iron garnet (Ce:YIG) is a widely known material because of its important magneto-optical (MO) properties over a wide range of wavelength spectrum (infrared to near UV-visible) \cite{onbasli2016optical,bi2011chip,MO_FOM,goto2013vacuum}. The strong MO activity and high transparency usually promotes Ce:YIG as a promising candidate for non-reciprocal device applications \cite{bi2011chip,shoji_magneto,stadler_integrated,zhang_dyCe}. Therefore, most of the studies were being focused for improving MO properties, such as Faraday and Kerr rotation in polycrystalline and/or epitaxial Ce:YIG films \cite{kim2001giant,kehlberger2015enhanced}. However, another interesting aspect in these kind of films is the development of perpendicular magnetic anisotropy (PMA). The fact that the garnet films are insulating in nature and exhibit low magnetic losses, the development of garnet films with PMA has opened immense applications in the fields of spintronics, magneonic,  spin caloritronics etc \cite{soumah2018ultra,avci2021current,wu_magnonValve,magnon_spintronic,uchida_SpinSeebeck}. For example, in pure and substituted YIG pure spin currents are observed due to the inverse spin Hall effect in a heavy metal over layer \cite{InverseSpinhall}, an electrical control of magnetization has also been realized in Tm\textsubscript{3}Fe\textsubscript{5}O\textsubscript{12} (TmIG) with adjacent heavy metal layer due to spin-orbit-torque (SOT) induced effect \cite{OnurAvci_Nature} etc. 

Recently we have studied epitaxial Ce:YIG films in reflection mode with longitudinal magneto-optical Kerr effect (MOKE) and shown that one can tune the MO activity by preparing the films in different $O_2$ partial pressure (OPP) during deposition and also by varying the thickness (deposition time) at a given OPP \cite{Manik_JAP}. However, we could not observe PMA in these films. In fact, there seems to be only one report dealing with the PMA in Ce:YIG films in literature, wherein the PMA is reported to be stable only at temperatures below 150 K \cite{CeYIG_PMA}. In view of this, in the present work we have undertaken the development of PMA in Ce:YIG films by further optimizing the deposition conditions with pulsed laser deposition (PLD), especially the substrate temperature during deposition and the substrate orientation. Moreover, it is observed that the PMA is stable over a wide temperature range ($\leq$400 K) enabling the possibility of room temperature based applications of these materials.  

\section{Experimental Details}

Pulsed laser deposition (PLD) technique is used to ablate stoichiometric Ce\textsubscript{1}Y\textsubscript{2}Fe\textsubscript{5}O\textsubscript{12} (Ce:YIG) target, the same target that has been used in our earlier work dealing with the Ce:YIG films prepared at different OPP \cite{Manik_JAP}. The films are grown in optimized O\textsubscript{2} partial pressure of about 110 mTorr and fixed substrate temperature of 780\textsuperscript{o}C. The other parameters i.e., laser influence, repetition rate and target to substrate distance are 2.5 J/cm\textsuperscript{2}, 6 Hz, 5 cm, respectively. All these conditions are kept same for every deposition except deposition time to vary the thickness. Three films of about 10, 30 and 65 nm thickness are prepared on Gd\textsubscript{3}Ga\textsubscript{5}O\textsubscript{12} (GGG) substrates. For every deposition, both the (100) and (111) oriented GGG substrates are mounted side by side, so that the films are deposited in same conditions.  The crystalline structure of the films are analyzed using x-ray diffraction (XRD) measurements carried out at angle dispersive X-ray diffraction (ADXRD) beamline (BL-12) of Indus-2, RRCAT India. The x-ray reflectivity (XRR) and reciprocal space mapping (RSM) are carried out using Bruker D8 diffractometer equipped with LynxEye detecotr, Goebel mirror and Eulerian cradle. Further surface roughness are determined by atomic force microscopy (AFM) measurements. 
Domain structure and magnetic hysteresis loops of the films are simultaneously captured with Kerr microscopy system of M/s Evico magnetics, Germany equipped with white light LED source. Room temperature ferromagnetic resonance (FMR) measurements are carried out using JEOL-FA200 Electron Spin Resonance Spectrometer. FMR spectra have been recorded as a function of polar angle ($\theta_H$) \cite{Binoy_FMR}. 

\begin{table}[t]
\small
\caption{\label{arttype} Notations of the prepared films and the obtained structural parameters. For example, Ce:YIG10-111 stands for the film of about 10 nm thickness and deposited on GGG(111) substrate. All the films are prepared at the same conditions except the deposition time i.e., film thickness. 't' is the thickness ($\pm$ 0.1 nm), $\rho$ is the roughness ($\pm$ 0.1 nm). $a_{out}$ is the out-of-plane lattice parameter ($\pm$ 0.0001 nm) obtained from Fig.~\ref{fig:XRD}. }
\begin{ruledtabular}
\begin{tabular}{ccccc}
Notation   & Substrate  & t   		& $\rho$  &$a_{out}$	 \\
             &   		 & (nm)    & (nm) & (nm)     \\
\hline 
Ce:YIG10-111  &GGG(111)  & 10.0    & 0.5  & 1.2383         \\
Ce:YIG30-111  &GGG(111)  & 30.0    & 0.9  & 1.2364  \\
Ce:YIG65-111  &GGG(111)  & 65.7    & 1.0  & 1.2350  \\
Ce:YIG10-100  &GGG(100)  & 9.2     & 0.5  & -       \\
Ce:YIG30-100  &GGG(100)  & 29.9    & 0.9  & -       \\
Ce:YIG65-100  &GGG(100)  & 64.0    & 0.2  & -       \\

\end{tabular}
\end{ruledtabular}
\label{tab:parameters}
\end{table}
 
\begin{figure}[htbp]
\centering
\includegraphics[height=5 cm, width=8 cm, keepaspectratio]{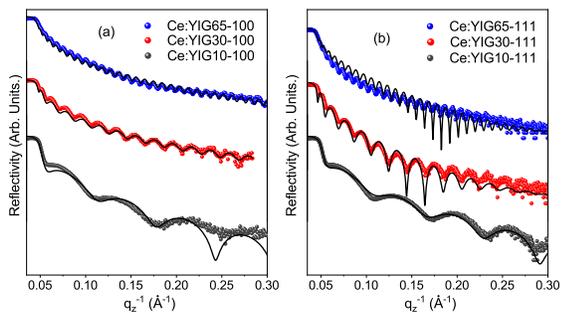}
\caption{X-ray reflectivity (XRR) data of the indicated films. Symbols represent the experimental data points and the solid line is the best fit to the data. Obtained parameters are shown in Table-1.}
\label{fig:XRR}
\end{figure}

\begin{figure}[htbp]
\centering
\includegraphics[width=8 cm]{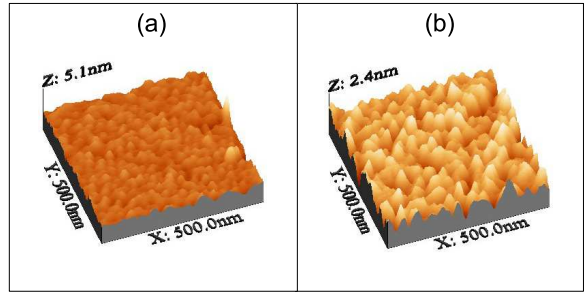}
\caption{Atomic force microscopy (AFM) images of (a) Ce:YIG10-111 (b) Ce:YIG66-111 films.}
\label{fig:AFM}
\end{figure}

\begin{figure}[b]
\centering
\includegraphics[width=8 cm]{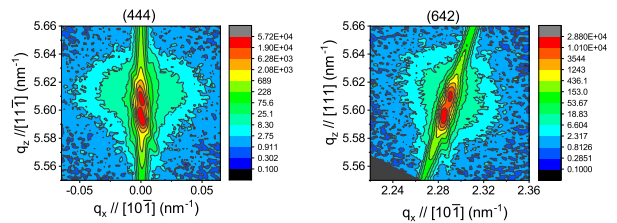}
\caption{Reciprocal space mapping (RSM) data of Ce:YIG66-111 measured across the indicated reflections. The two intense peaks (red color) are substrate $\alpha_1$ and $\alpha_2$ reflections.}
\label{fig:RSM}
\end{figure}

\begin{figure*}[htbp]
\centering
\includegraphics[width=\textwidth]{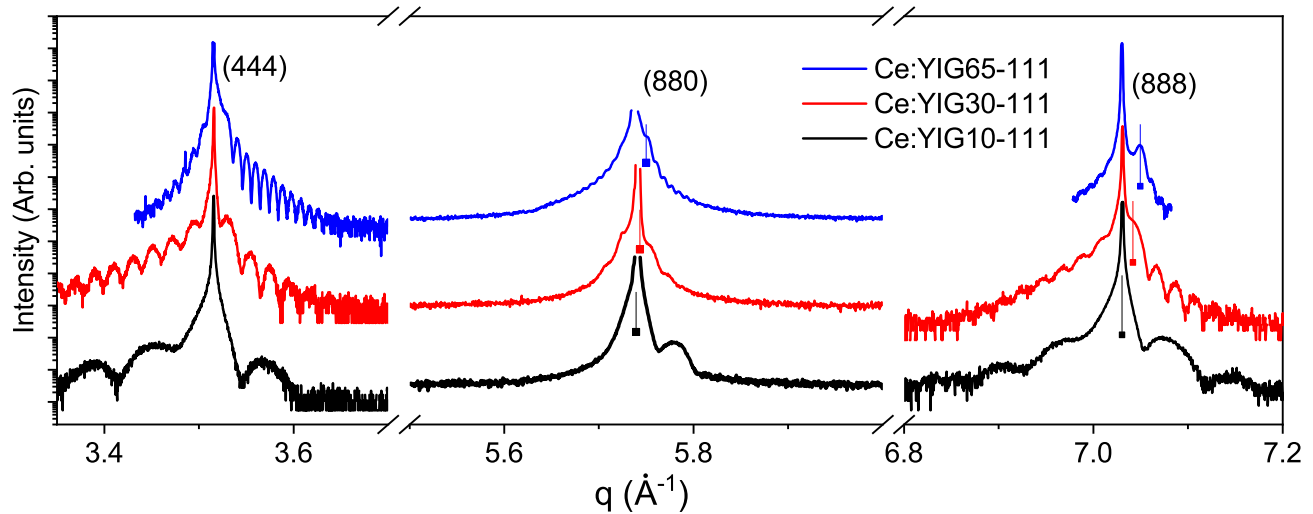}
\caption{2$\theta$-$\omega$ scans of the Ce:YIG films deposited on GGG(111) substrate measured with synchrotron radiation. The data is measured across symmetric (444), (888) reflections \& asymmetric (880) reflection. The small vertical lines are to indicate the film peak.}
\label{fig:XRD}
\end{figure*}
  
\section{Results}

Fig.~\ref{fig:XRR} show XRR patterns of Ce:YIG films deposited on GGG(111) and GGG(100) substrates. Fitting to the experimental data using Parratt formalism gives the parameter such as thickness,   roughness and the obtained parameters are shown in table-I. Further, the surface morphology of the films is studied with atomic force microscopy and Fig.~\ref{fig:AFM} shows the AFM images 10 and 66 nm thick films deposited on GGG(111) substrate. The data indicate a smooth surface with $\textit{rms}$ roughness less than 1 nm, corroborating the XRR data.  

Since the lattice parameter of the Ce:YIG films deposited at higher OPP and substrate temperature is expected to be very close to that of GGG substrate\cite{Manik_JAP}, the film peaks are merged with that of substrate in the 2$\theta$-$\omega$ scans (not shown) of all the films measured using lab based x-ray diffractometer. However no other reflections are observed in the 2$\theta$-$\omega$ scans indicating that there are no secondary phases present for all the samples. This is consistent with our earlier work in which the film Bragg peaks are observed to shift close to GGG substrate peak at 110 mTorr OPP \cite{Manik_JAP}.  Further, reciprocal space mapping (RSM) data across symmetric and asymmetric reflections are carried out and the representative data of Ce:YIG66-111 film across (444) and (642) reflections is shown in Fig.~\ref{fig:RSM}. Since the present RSM measurements are not done in high-resolution mode with lab based x-ray diffractometer (as a result one can see the substrate $\alpha_2$ reflection also) the film and substrate reciprocal lattice points are overlapping with each other, even though the data confirms that the films are epitaxial in nature. However, for samples deposited on GGG(111) substrates (which exhibit PMA as discussed below) the 2$\theta$-$\omega$ scans are carried out using synchrotron radiation and the data is shown in Fig.~\ref{fig:XRD}.

Fig.~\ref{fig:XRD} shows the XRD measured around symmetric (444), (888) \& asymmetric (880) reflections of all the films deposited on GGG(111) substrate. All the films exhibit pronounced Laue oscillations which clearly indicate that the films are highly smooth, uniform and single crystalline in nature. It may be noted that the thickness of the films calculated from Laue oscillations match excellently with the XRR data. The out-of-plane (OOP) lattice parameter ($a_{out}$) is estimated from the (888) reflection data and is tabulated in table-1. Usually one may expect that the film peak shifts to higher (lower) angles with respect to the substrate, below a critical thickness of the film, due to the expected in-plane tensile (compressive) epitaxial strains in the case of lattice mismatch between substrate and film \cite{TbandEuIG_strain}. However, it is observed that for 10 nm thick sample, the film peak is overlapping with that of substrate, which could be due to the excellent lattice matching between the film and substrate. With further increase in thickness, the OOP lattice parameter is observed to decrease indicating the OOP lattice compression, unlike most of the cases in which films are expected to relax as thickness increases. The plausible reason for this could be either due to changes in the relative concentration of Ce\textsuperscript{3+}, Ce\textsuperscript{4+} and associated issues related to charge neutralization in the system with thickness \cite{Manik_JAP, ExcessOxygen}.

\begin{figure}[htbp]
\centering
\includegraphics[width=8 cm]{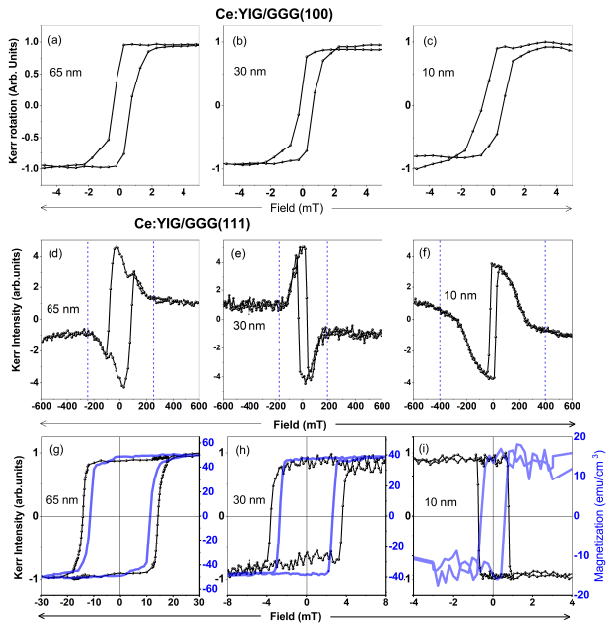}
\caption{(a)-(c) L-MOKE loops of Ce:YIG films deposited on GGG(100) substrate. (d)-(f) L-MOKE loops of Ce:YIG films deposited on GGG(111) substrate. The vertical dotted lines of L-MOKE loops is to indicate the in-plane saturation field (H\textsubscript{Sat}). (g)-(i) P-MOKE (black color) and SQUID-VSM (blue color) loops of Ce:YIG films deposited on GGG(111) substrate. Thickness of the films is indicated in the respective frame. }
\label{fig:RTMOKE_SQUID}
\end{figure}

Fig.~\ref{fig:RTMOKE_SQUID} show the MOKE data of all the films along with the bulk magnetization data measured using SQUID-VSM for few selected samples. MOKE measurements are carried out in two geometries viz., longitudinal (L-MOKE) in which the applied magnetic field is in the film plane and polar (P-MOKE) in which the applied field is out-of-plane of the sample. P-MOKE measurements are sensitive to out-of-plane component of the magnetization and the L-MOKE is expected to be sensitive to in-plane component of magnetization \cite{Rudi_book2}. However, it may be noted that L-MOKE geometry can also have a sensitivity of out-of-plane magnetization component in addition to the in-plane component due to the oblique incidence of light \cite{Rudi_book2}. Perusal of Fig.~\ref{fig:RTMOKE_SQUID}, for all the films deposited on GGG(100) substrate the observed L-MOKE data is typical of films exhibiting in-plane magnetization.  

However, for the films deposited on GGG(111) substrates, an anomalous kind of L-MOKE loops are observed as shown in Fig.~\ref{fig:RTMOKE_SQUID}. The data indicates two switching fields (Fig.~\ref{fig:RTMOKE_SQUID}(d),(e),(f)). As one can see that at middle portion of the L-MOKE loops, a clear switching of magnetization with finite H\textsubscript{C} is observed in addition to the saturation at sufficiently high magnetic fields as indicated by vertical dashed line in Fig.~\ref{fig:RTMOKE_SQUID}(d),(e),(f), the magnetization saturates along the field direction. The central switching part could be due to the presence of considerable OOP magnetization component or domains with non-zero OOP field components during application of in-plane external magnetic field. In view of this, P-MOKE measurements are also carried out on films deposited on GGG(111) substrate. Fig.~\ref{fig:RTMOKE_SQUID} shows the room temperature P-MOKE hysteresis loops. It is very interesting to note that the observed squareness (ratio of remanence/saturation, M\textsubscript{R}/M\textsubscript{S}) is close to unity for all the three films in the P-MOKE data. In order to confirm this, bulk magnetization measurements with SQUID-VSM are also carried out on these three films with the applied field along the out-of-plane direction. The SQUID-VSM data is observed to match excellently with P-MOKE data except small variation in the coercivity (H\textsubscript{C}). Therefore, both P-MOKE and SQUID-VSM data clearly suggest that at room temperature the easy magnetization axis is found to be along out-of-plane direction for the films deposited on GGG(111) substrate. 

\begin{figure}[htbp]
\centering
\includegraphics[width=8.5 cm]{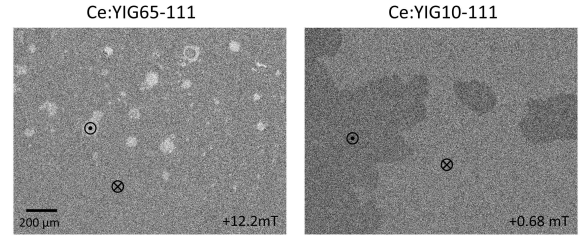}
\caption{Room temperature Kerr micrograph of the indicated films captured in P-MOKE geometry. Scale bar is same for both the images. $\bigodot$ and $\bigotimes$ denote out-of-plane (OOP) and in to the plane respectively.}
\label{fig:RT_Domain}
\end{figure}

Development of PMA is further substantiated with magnetic microstructure of the films by capturing Kerr images. It may be noted that when the preferred (easy) direction of magnetization is perpendicular to the plane of the film, usually magnetic bubble domains are expected to form \cite{Rudi_Book1}. Bubble domains are cylindrical magnetic domains that may occur in a thin plate of a magnetic material exhibiting PMA. Fig.~\ref{fig:RT_Domain} shows room temperature Kerr images of Ce:YIG10-111 and Ce:YIG66-111 films captured in P-MOKE geometry. The samples are initially saturated by applying a sufficient negative field and the images are captured at the indicated positive field, which is close to H\textsubscript{C} and one can clearly see the bubble domain formation confirming the presence of PMA in the Ce:YIG films deposited on GGG(111) substrate. 
In order to further confirm the PMA, the FMR measurements are carried out on two films of each that are deposited on GGG(100) and GGG(111) substrates and the results are discussed as following.  

\begin{figure}[htbp]
\centering
\includegraphics[width=8.5 cm]{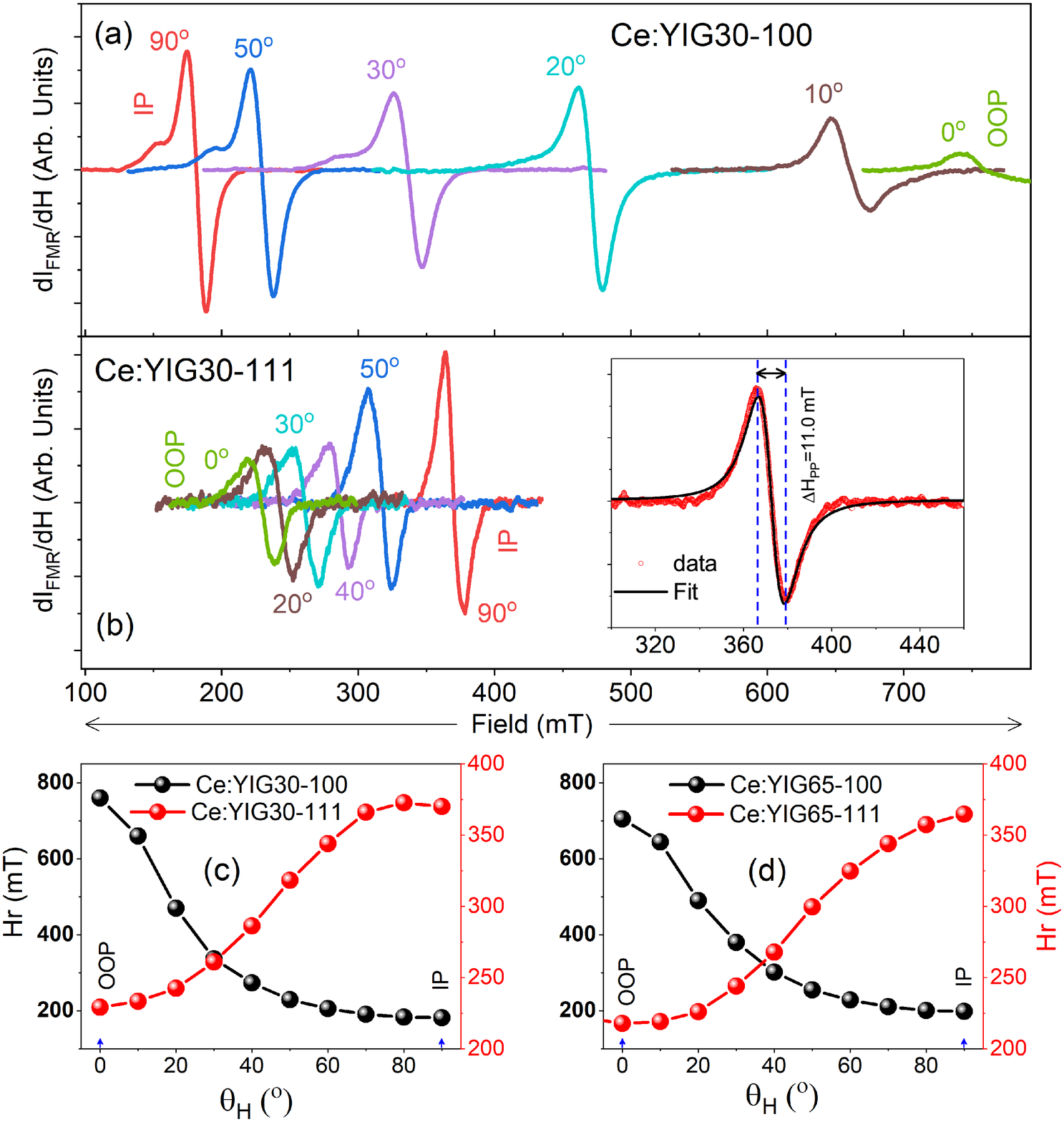}
\caption{(a),(b) Ferromagnetic resonance (FMR) data of the indicated films. Inset of (b) shows the fitting of one representative FMR data to extract resonance field (H\textsubscript{r}) and typical line width ($\Delta$$H_{pp}$). IP and OOP denote in-plane and out-of-plane configuration respectively. (c),(d) show the variation of H\textsubscript{r} with OOP angle ($\theta_H$), the solid line is guide to eye only.}
\label{fig:FMR}
\end{figure}

Fig.~\ref{fig:FMR} shows the FMR derivative spectra of the indicated films measured as a function of polar angle, $\theta_{H}$, between the normal of the film plane and the applied field. FMR signal for 10 nm thick films was weak and hence is not shown. The resonance field values ($H_{r}$) extracted from the FMR data as a function of $\theta_{H}$ is also shown in Fig.~\ref{fig:FMR}. One can clearly see from the data that for the films deposited on GGG(111) substrate the resonance field (H\textsubscript{r}) values are the smallest in OOP configuration ($\theta_{H}$=$0^{\circ}$) and the largest in IP configuration ($\theta_{H}$=$90^{\circ}$) whereas for films deposited on GGG(100) substrate, the trend is opposite. This indicates that easy axis magnetization is out of plane (i.e., PMA) for the films deposited on GGG(111) substrate and it is in-plane for the films deposited on GGG(100) substrate. 
The observed peak to peak lines width ($\Delta$$H_{pp}$) for studied samples are found to be larger as compared to the pure YIG films \cite{YIG_PMA,FMR_YIG1,FMR_YIG2}. This indicate in the studied Ce:YIG samples have higher magnetic damping as also expected due to lattice dilation and different environments of the Fe cations due to the Ce doping \cite{kehlberger2015enhanced,vasili2017direct}. It is also to be mentioned that spin orbit coupling present in the rare earth iron garnet gives rise to larger magnetic damping than pure YIG. However, the magnetic damping in our Ce:YIG samples is found to be comparable or lesser than TmIG and other rare earth garnet those are showing PMA \cite{Damping_TmIG1,Damping_TmIG1,Damping_EuIG}.   

\begin{table}[b]
\small
\caption{\label{arttype} M\textsubscript{s} is the saturation magnetization obtained from SQUID measurements. H\textsubscript{K} and K\textsubscript{U} denote out-of-plane (OOP) effective anisotropy field and anisotropy energy density, respectively. K\textsubscript{ME} is the magneto elastic anisotropy density.  H\textsubscript{C(PMA)} is the coercivity from the P-MOKE data (Fig.~\ref{fig:RTMOKE_SQUID})}, 
\begin{ruledtabular}
\begin{tabular}{cccccc}
Sample    & 4$\pi$M\textsubscript{S} & H\textsubscript{K}	& K\textsubscript{U}  & K\textsubscript{ME} & H\textsubscript{C(PMA)}   \\
					& (Gauss)      							& (Oe)      				& ($kJ/m^{3}$)  			& ($kJ/m^{3}$) 				& (mT)    \\
\hline 
Ce:YIG10-111  & 200$\pm$50  & 4200$\pm$200 & 3.4   & -        & 0.8   \\
Ce:YIG30-111  & 500$\pm$25  & 2500$\pm$50  & 5.0   & 0.50    & 3.6   \\
Ce:YIG65-111  & 630$\pm$12  & 3130$\pm$40  & 7.8   & 0.82   & 14.5 \\

\end{tabular}
\end{ruledtabular}
\label{tab:magparameters}
\end{table}

The above observations clearly indicate that the PMA is developed in films that are deposited on GGG(111) substrates, whereas the films deposited on GGG(100) substrate exhibit only in-plane magnetization. One can understand this by considering various magnetic anisotropy terms and their origin such as intrinsic magneto-crystalline anisotropy (K\textsubscript{MC}), magneto-elastic (K\textsubscript{ME}), shape anisotropy (K\textsubscript{Shape}) and induced anisotropy as discussed in the following. 

In general in YIG and most of the other RIG systems, because of negative cubic anisotropy constant (K\textsubscript{1}), easy axis of magnetization is always along $\langle 111 \rangle$ direction \cite{MCandME_Anisotropy}, which is an inherent magneto-crystalline anisotropy factor (K\textsubscript{MC}). In addition to this one can also understand the easy axis of magnetization as $\langle 111 \rangle$ in terms of magneto-striction coefficient ($\lambda$) i.e., magneto-elastic anisotropy (K\textsubscript{ME}). Depending upon $\lambda$, either compressive or tensile strain due to lattice mismatch between film and substrate can produce large uniaxial magneto-elastic anisotropy to overcome shape anisotropy for inducing PMA\cite{Rashba_TmIG,StrainTune_YIG,TbandEuIG_strain}. For example, considering YIG, the $\lambda$ values are given by $\lambda_{111}$=-2.73 $\times$ $10^{-6}$ and $\lambda_{100}$=-1.25 $\times$ $10^-6$, which are usually small and negative \cite{MCandME_Anisotropy}. However, larger in magnitude of $\lambda_{111}$ than $\lambda_{100}$ make $\langle 111 \rangle$ direction as a favorable easy axis than $\langle 100 \rangle$, when a compressive stress is applied along the respective direction.  Along these lines, it may be noted that from our XRD data, the presence of in-plane tensile strain is argued for higher thickness films which require a negative $\lambda$ value to generate PMA in the present samples. However, it is reported in literature that the Ce substitution in YIG makes $\lambda$ from negative to positive with increasing Ce concentration \cite{CeYIG_Magnetostriction1,CeYIG_Magnetostriction2} indicating that in addition to this magneto-elastic contribution there could be other contribution for the observed PMA in the present work. For this we estimated quantitatively magneto-elastic anisotropy (K\textsubscript{ME}) as discussed in the following.

The effective PMA field (H\textsubscript{K}) values can be calculated using equation H\textsubscript{Sat} = H\textsubscript{K}-4$\pi$M\textsubscript{S}, where H\textsubscript{Sat} is IP saturation field which is obtained from L-MOKE loops and M\textsubscript{S} is the saturation magnetization deduced from SQUID-VSM M-H data (Fig.~\ref{fig:RTMOKE_SQUID}).  The anisotropy K\textsubscript{U} is related through the equation K\textsubscript{U}= (H\textsubscript{K}$M\textsubscript{S})/$2 \cite{YIG_PMA}. The observed value of K\textsubscript{U} of the three studied PMA films are tabulated in the table-II. 

The K\textsubscript{ME} is calculated for the Ce:YIG30-111 and Ce:YIG65-111 films using equation K\textsubscript{ME}=(9/4)$\lambda_{111}$ $C_{44}$ ($\pi/2$-$\beta$), where $C_{44}$ is the shear stiffness constant and $\beta$ is the shear distortion angle. The $\beta$\cite{Tunable_ME} can be deduced from IP and OOP lattice parameter values as obtained from x-ray diffraction data (Fig.~\ref{fig:XRD}), where the IP lattice parameter of the film is considered to be equivalent to that of substrate due to coherent growth. The observed $\beta$ are found to be 90$^{\circ}$ (i.e., no distortion), 90.06$^{\circ}$ and 90.1$^{\circ}$ for Ce:YIG10-111,Ce:YIG30-111 and Ce:YIG65-111, respectively. Substitution of thus obtained $\beta$, $\lambda_{111}$=-2.73 $\times$ $10^{-6}$ and $C_{44}$=76.6 GPa  results in K\textsubscript{ME} as 0.50$\pm$0.02, and 0.82$\pm$0.02, $kJ/m^{3}$ for Ce:YIG30-111 and CeYIG65-111 films, respectively. We considered $\lambda_{111}$ and $C_{44}$ values of YIG \cite{YIG_PMA,ME_YIG} in the above calculation mainly due to the unavailability of these values for Ce:YIG samples, but this calculation is expected to give an upper limit for the values of values of K\textsubscript{ME}. 

\begin{figure}[htbp]
\centering
\includegraphics[width=8.5 cm]{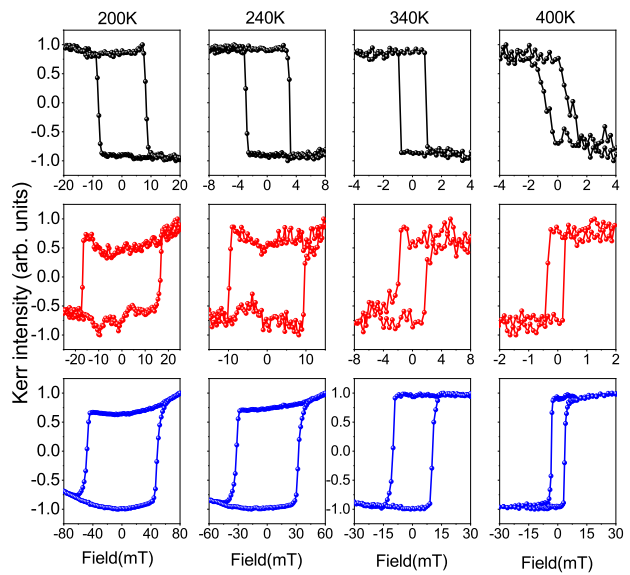}
\caption{Temperature dependent P-MOKE data of Ce:YIG films deposited on GGG(111) substrate (top) 10 nm thickness (middle) 30 nm thickness (bottom) 66 nm thickness at the indicated temperatures. }
\label{fig:PMOKE_T}
\end{figure}

Clearly, experimentally observed OPP anisotropy K\textsubscript{U} (ref table-II) is one order of magnitude large than what is obtained from respective K\textsubscript{ME}. Even the magneto-elastic field H\textsubscript{ME}, defined as H\textsubscript{ME}=2K\textsubscript{ME}/M\textsubscript{S} is not even 50$\%$ of the demagnetizing fields (4$\pi$M\textsubscript{S}) of the respective films and therefore K\textsubscript{ME} cannot be solely responsible for the observed PMA. An extra anisotropy term is therefore required to be introduced, which is often called growth induced anisotropy term (K\textsubscript{GROWTH}) as reported by Soumah et al., in Bi substituted YIG films \cite{soumah2018ultra}. In view of the small value of K\textsubscript{MC}, K\textsubscript{U} is attributed to be sum of two contributions i.e., K\textsubscript{ME} and K\textsubscript{GROWTH} for Ce:YIG30-111 and Ce:YIG30-111 films whereas in case of Ce:YIG10-111 film, due to absence of strain, only K\textsubscript{GROWTH} term is expected to be responsible for the observed PMA. However the exact microscopic origin of K\textsubscript{GROWTH} and its dependence on factors such as composition of the film, thickness, orientation etc., is beyond the scope of the present work.

\begin{figure}[htbp]
\centering
\includegraphics[width=8.5 cm]{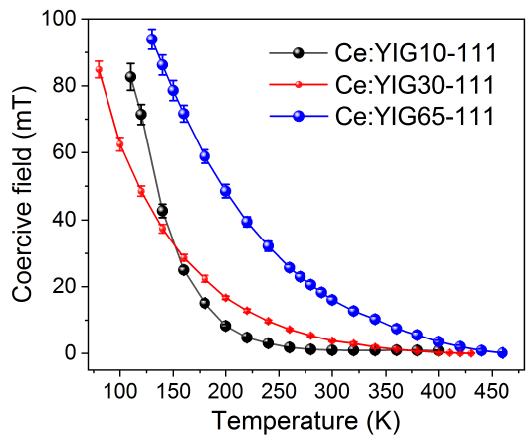}
\caption{Temperature variation of coercivity (H\textsubscript{C}) obtained from Fig.~\ref{fig:PMOKE_T} of all the Ce:YIG films deposited on GGG(111) substrate. }
\label{fig:HC}
\end{figure}

\begin{figure}[htbp]
\centering
\includegraphics[width=8.5 cm]{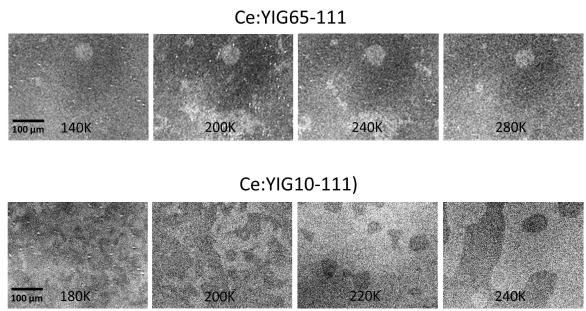}
\caption{Kerr micrograph of Ce:YIG films at the indicated temperatures captured in P-MOKE geometry. Scale bar is same for all the images.}
\label{fig:Domain_Temp}
\end{figure}

The stability of the developed PMA in Ce:YIG films deposited on GGG(111) is further studied with temperature. Temperature dependent P-MOKE measurements are carried out over a temperature range of 80 to 460 K and the data for all the three films is shown in Fig.~\ref{fig:PMOKE_T}. It is observed that the easy axis of magnetization remain perpendicular to the film plane irrespective of film thicknesses much above the room temperature as shown in Fig.~\ref{fig:PMOKE_T}. The apparent curved shape of the loops is due to the Faraday contribution of the objective lens of the microscope at higher fields. This is a remarkable stability of the PMA in these kind of films as very few reports could show this kind of robust stability of PMA in garnet epitaxial thin films.  For example, reorientation from out-of-plane to in-plane easy axis at about 173 K is reported in similar Ce:YIG epitaxial films, prepared at different growth conditions \cite{CeYIG_PMA}. 

The coercivity (H\textsubscript{C}) is estimated from the data of Fig.~\ref{fig:PMOKE_T} for all the films and the variation of H\textsubscript{C} is shown in Fig.~\ref{fig:HC}. The increase of H\textsubscript{C} with decreasing temperature is observed for all the films as expected. The temperature dependent Kerr images are captured near H\textsubscript{C} during the high temperature P-MOKE measurements and the data is shown in Fig.~\ref{fig:Domain_Temp}. For Ce:YIG65-111 film, the bubble nucleation point and the size of the the bubble is found to be almost same irrespective of temperature.  However for lower thickness film (Ce:YIG10-111) the size of magnetic bubble is found to increase with temperature.

\maketitle \section{Conclusions}

In conclusion, the present work reports the stabilization of perpendicular magnetic anisotropy (PMA) in cerium substituted yttrium iron garnet (Ce:YIG) epitaxial thin films on GGG substrate by suitably optimizing the deposition conditions with pulsed laser deposition. The developed PMA is confirmed from bulk and surface sensitive magnetic measurements, which is further corroborated by ferromagnetic resonance and magnetic microstructure. The observed stability of PMA over a wide range of temperatures makes the studied Ce:YIG films suitable for many practical applications. The cerium substitution in YIG seems to be a better choice for the insulating garnet films as the possibility of enhancing magneto-optical activity\cite{Manik_JAP} and stabilization of PMA are demonstrated with suitably optimized deposition conditions.  

\maketitle \section{Acknowledgments}
Dr.R.Venkatesh and Mr. Mohan Gangrade are thanked for the AFM measurements. MK thank Dr.Zaineb Hussain for the discussions and Mr.Rakhul Raj for the help during MOKE measurements. VRR thank Prof. Ajay Gupta for the discussions and encouragement. 

\maketitle \section{References}

\bibliography{Reference_Garnet1}

\end{document}